\documentclass[preprint,prl,aps,amsmath,amssymb,color,superscriptaddress]{revtex4}
\usepackage{stmaryrd}
\usepackage{amsmath}
\usepackage{amssymb}
\usepackage{graphicx}
\usepackage{dcolumn}
\usepackage{bm}
\usepackage{multirow}
\usepackage{tabularx}
\usepackage{epsfig}
\usepackage{url}

\begin{document}
\begin{titlepage}
\title{Stable phases of freestanding monolayer TiO$_2$: The emergence of out-of-plane ferroelectricity}
\author{Xiangtian Bu}
\affiliation {Key Lab of advanced optoelectronic quantum architecture and measurement (MOE), and Advanced Research Institute of Multidisciplinary Science, Beijing Institute of Technology, Beijing 100081, China}
\author{Haitao Liu}
\affiliation {Institute of Applied Physics and Computational Mathematics, Beijing 100088, China}
\affiliation {National Key Laboratory of Computational Physics, Beijing 100088, China}
\author{Yuanchang Li}
\email{yuancli@bit.edu.cn}
\affiliation {Key Lab of advanced optoelectronic quantum architecture and measurement (MOE), and Advanced Research Institute of Multidisciplinary Science, Beijing Institute of Technology, Beijing 100081, China}

\date{\today}

\begin{abstract}
Despite being successfully synthesized [Zhang $et$ $al.$, Nat. Mater. \textbf{20}, 1073 (2021)], the monolayer structure of stable hexagonal TiO$_2$ is unknown, and it is not even clear whether it can exist in a freestanding form. Through first-principles calculations, we have identified two previously uncharted stable structures, namely, distorted 1$\times$$\sqrt{3}$ 1T-TiO$_2$ and $\sqrt{3}$$\times$$\sqrt{3}$ 1T-TiO$_2$, both of which are energetically more favourable than commonly adopted 1H and 1T phases. Here structural distortions are characterized by the out-of-plane shifts of Ti atoms due to the pseudo-Jahn-Teller interactions, which break one and all two inversion symmetries of 1T configuration. As a consequence, the 1$\times$$\sqrt{3}$ 1T remains centrosymmetric while the $\sqrt{3}$$\times$$\sqrt{3}$ 1T exhibits out-of-plane ferroelectricity. Electronic structure calculations show that both two are wide-bandgap semiconductors with bandgaps larger than their bulk counterparts. Our study not only deepens the understanding of structural instability in wide-gap semiconductors but also adds a new member to the rare family of two-dimensional out-of-plane ferroelectrics.
\end{abstract}

\maketitle
\draft
\vspace{2mm}
\end{titlepage}

Titanium dioxide (TiO$_2$) has long been recognized as one of the most promising photocatalysts for the production of hydrogen from water using solar energy\cite{nature1972}. Its application, however, has been hampered by a wide bandgap of more than 3 eV, which limits it to absorbing only the ultraviolet portion of the solar spectrum. As this fraction accounts for only $\sim$5\% of the solar radiation on Earth, efforts have been made to narrow the bandgap of TiO$_2$ in order to utilize a wider range of the solar spectrum and improve the conversion efficiency. Current bandgap optimization strategies focus on introducing mid-gap states through doping\cite{Asahi,Gai,Yin,Zhu,Wang}, but this has proven to be difficult. Two-dimensional TiO$_2$ offers an alternative as its dopant-free phase has been found to have a narrow bandgap $\sim$2.1 eV\cite{NC2011}. Recently, a planar hexagonal TiO$_2$ monolayer is synthesised by strictly controlled oxidation at the metal-gas interface with a bandgap of 2.35 eV\cite{NM2021}, which is almost optimal for visible light absorption. However, its unknown crystal structure constrains the understanding of the bandgap reduction.

Generally, the bandgap of layered materials increases as the number of layers decreases due to quantum size effects. For non-layered materials, such as III-V semiconductors, one does find a bandgap decrease for monolayers, but the crystal structure changes significantly compared to the bulk phase\cite{Jiang,Dong,Qin}. Regarding the hexagonal TiO$_2$ monolayer, it is still not even clear whether it can be stabilized in a freestanding form. Although the 1H-TiO$_2$ bandgap calculated using the Heyd-Scuseria-Ernzerhof functional is in agreement with the experimental results\cite{NM2021}, this alone is not sufficient to determine the structure, because of the well-known fact that the bandgap of transition-metal oxides from DFT-based approaches depends strongly on the exchange-correlation functionals used\cite{Ataca,PRM2020,Rasmussen,TanJCP}. In terms of formation energy, 1H-TiO$_2$ is 0.36 (0.53) eV/atom higher than 1T-TiO$_2$ by the PBE (PBE + $U$) calculation\cite{Rasmussen,PRM2020}, suggesting a difficulty in synthesizing it at thermal equilibrium. More importantly, both 1H- and 1T-TiO$_2$ are unstable with imaginary frequencies on their phonon spectra\cite{Ataca}. It has been reported that the imaginary frequencies of 1T-TiO$_2$ can be eliminated when considering Hubbard $U$\cite{PRM2020}. Nevertheless, this elimination depends on the size of $U$ and there is no universal $U$ that reproduces the crystal and electronic structure, as well as the phase stability, of TiO$_2$ at the same time\cite{Dompablo}. Historically, a stable structure known as Lepidocrocite-type has been proposed\cite{jpcb2003}, but its out-of-plane O-Ti-O-O-Ti-O layer order renders it more like a bilayer. In addition, the Ti atoms therein form a tetragonal rather than a hexagonal lattice.

In this work, we perform extensive first-principles calculations on the structural and electronic properties of hexagonal TiO$_2$ monolayer and identify two stable configurations, namely, distorted 1$\times$$\sqrt{3}$ 1T-TiO$_2$ and $\sqrt{3}$$\times$$\sqrt{3}$ 1T-TiO$_2$ (denoted hereafter as $d$1T and $t$1T, respectively, depending on the number of Ti atoms in the unit cell). This distortion manifests itself as an out-of-plane shift of Ti, which on the one hand reduces the system energy and eliminates imaginary frequencies, thus stabilizing the structure. On the other hand, it breaks one and all of the two spatial inversion symmetries of 1T to form $d$1T and $t$1T, respectively. As a result, the motions of the two Ti atoms in $d$1T are always centrosymmetric, whereas in $t$1T the inversion symmetry is spontaneously broken, giving rise to an inherent out-of-plane ferroelectricity of 0.5 $\mu$C/cm$^2$. Unlike in transition-metal dichalcogenides where ferroelectricity appears across the metal-insulator transition\cite{Shirodkar,Bruyer}, 1T-TiO$_2$ is a wide-gap semiconductor whose structural instability is driven by the pseudo-Jahn-Teller mechanism\cite{pjet}. Electronic structure calculations show that bandgaps of both $d$1T and $t$1T are larger than that of bulk TiO$_2$, hinting to an important influence of the substrate on the experimentally observed bandgap narrowing\cite{NM2021}. Our study not only deepens the understanding of structural instability in wide-gap semiconductors but also adds a new member to the rare family of two-dimensional out-of-plane ferroelectrics.

All density-functional-theory (DFT) calculations were performed within the formalism of the Perdew-Burke-Ernzerhof (PBE) \cite{pbe} exchange-correlation functional as implemented in the Vienna ab initio simulation package (VASP)\cite{vasp}. The projector augmented-wave method\cite{paw} was employed with an energy cutoff of 520 eV. A vacuum layer of 18 \AA\ was used to minimize spurious interactions between adjacent images. Integration over the Brillouin zone was done using the Monkhorst-Pack $k$-point grids of 24 $\times$ 24 $\times$ 1, 24 $\times$ 15 $\times$ 1, 18 $\times$ 18 $\times$ 1 and 8 $\times$ 8 $\times$ 12 for 1H/1T, $d$1T, $t$1T monolayer and bulk rutile. The lattice constants and atomic positions were fully optimized until the residual force on each atom was less than 1 meV/\AA. The van der Waals interaction was included using the Grimme's D2 scheme\cite{vdw}, which is considered necessary to reproduce the fact that the bulk rutile is lower in energy than the anatase\cite{PRM2020,NM2021}. Quasiparticle $G_{0}W_{0}$ calculations\cite{gw} were performed to fix the bandgap problem of PBE, with the number of empty bands exceeding 6 times the valence bands. Optical gaps and exciton spectrum were obtained by solving the Bethe-Salpeter equation (BSE)\cite{bse} on top of the $G_{0}W_{0}$ results. Ferroelectric polarization was calculated using the Berry-phase method\cite{berry}. The phonon spectra were calculated within density functional perturbation theory using the Quantum ESPRESSO package\cite{Giannozzi_2017}, with the cutoff energy of 80 Ry, the same $k$-point grids as above, and the $q$-point grids of 8 $\times$ 8 $\times$ 1, 8 $\times$ 5 $\times$ 1 and 6 $\times$ 6 $\times$ 1 for 1H/1T, $d$1T and $t$1T.

\begin{figure*}[htbp]
\includegraphics[width=0.98\columnwidth]{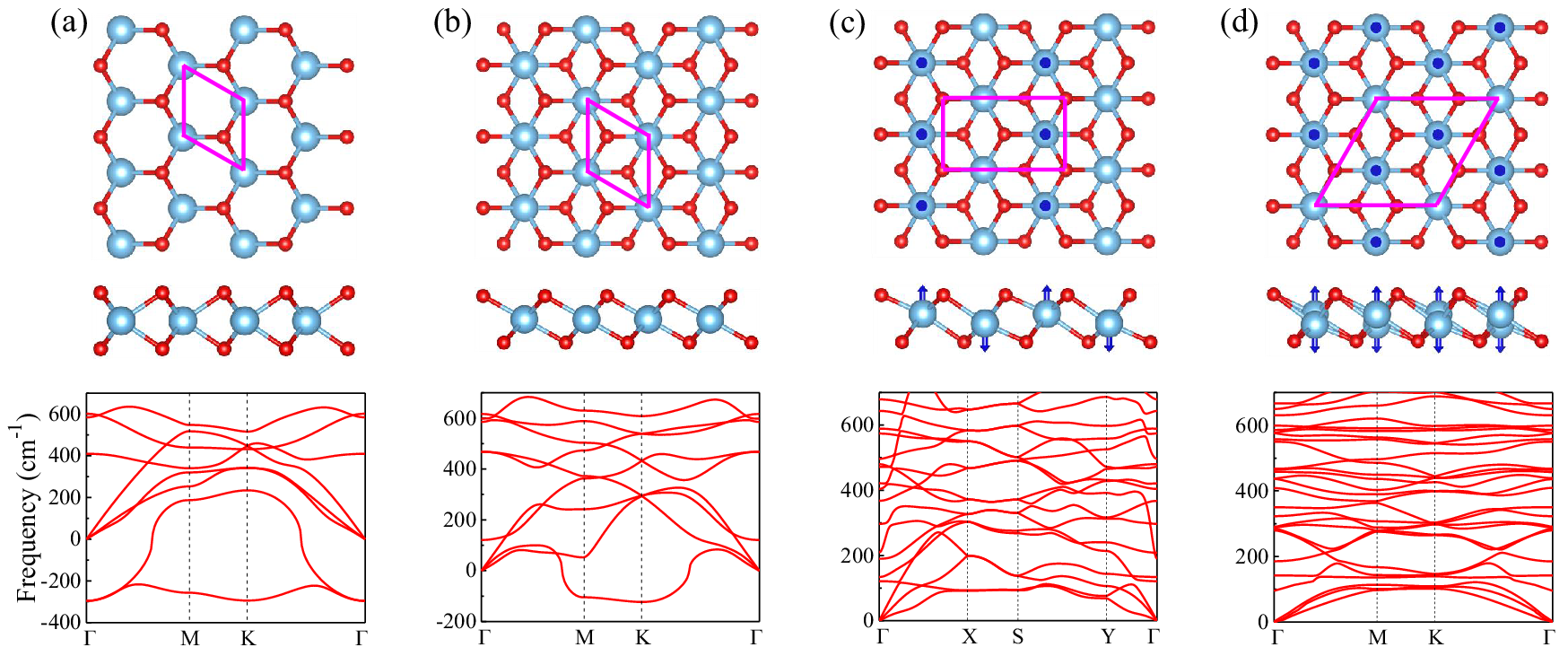}
\caption{\label{Fig1} (Color online) Geometric structures and phonon spectra of monolayer TiO$_2$ with the configuration of (a) 1H, (b) 1T, (c) $d$1T, and (d) $t$1T. First and second row corresponds to top and side views, respectively. Pink rectangles denote the unit cells, and cyan and red balls denote Ti and O atoms. In (c) and (d), blue arrows mark the Ti displacements relative to their positions in the 1T configuration.}
\end{figure*}

Honeycomb-like 1H and 1T structures are common in monolayers of transition-metal oxides and dichalcogenides. However, both 1H- and 1T-TiO$_2$ are unstable, as evidenced by the imaginary frequencies on their phonon spectra [see Figs. 1(a) and 1(b)]. The presence of a ``soft" phonon band throughout the Brillouin zone means that 1H is unlikely to exist in a freestanding form. In contrast, the soft phonon modes of 1T are predominantly present on the $MK$ line, which can usually be eliminated by enlarging the cell and allowing relative motion between otherwise equivalent atoms.

When relaxation starts from slightly deviating perfect 1$\times$$\sqrt{3}$ 1T and $\sqrt{3}$$\times$$\sqrt{3}$ 1T supercells, we obtain the stable structures $d$1T and $t$1T, as shown in Figs. 1(c) and 1(d), respectively. Both no longer have soft phonon modes. The energy of $d$1T is 19 meV/Ti lower than that of 1T, while the energy of $t$1T is further reduced by 7 meV/Ti. Note that in $d^2$ transition-metal dichalcogenides such as MoS$_2$, different phases (2H, 1T and 1T') are experimentally observed\cite{Kappera,Liuli}, even though their energy differences are an order of magnitude larger than here\cite{Duerloo}. It is therefore reasonable to speculate that both $d$1T and $t$1T could potentially be prepared experimentally.

We are aware that bulk TiO$_2$ has a variety of phases whose relative stability depends on the different exchange-correlation functionals used in the calculations\cite{Dompablo}. For this reason we have performed HSE06 and PBE + $U$ calculations. The HSE06 is consistent with the PBE in that both predict structural distortions. Specifically, the energy of $d$1T is 8 meV/Ti lower than that of 1T, while the energy of $t$1T is further reduced by 7 meV/Ti. For comparison, the PBE + $U$ results show a dependence on the value of $U$. Increasing $U$ tends to suppress the distortion by decreasing the energy gain of the structural distortion. At $U$ = 3 eV, the structural distortion disappears completely. Previous studies on bulk TiO$_2$\cite{Dompablo} have shown that there is no universal $U$ that can simultaneously provide the correct cell parameters, relative phase stability, and electronic gap. Furthermore, around $U$ = 3 eV, the PBE + $U$ fails to predict the structural instability of TiSe$_2$\cite{Hellgren}. In contrast, the HSE06 well reproduces the cell parameters and electronic gap of bulk TiO$_2$, as well as the structural instability of TiSe$_2$. However, it fails to reproduce the relative stability of rutile and anatase TiO$_2$. Although more experimental evidence is needed to test whether nonlocal exchange or on-site electron-electron interactions dominate in TiO$_2$, there is no doubt that its hexagonal monolayer is indeed capable of existing in the freestanding form.

Here $d$1T and $t$1T essentially maintain the in-plane hexagonal skeleton, with the most pronounced deformation being the out-of-plane shifts of the Ti atoms. Specifically, the two Ti in $d$1T are symmetrically shifted upward and downward by 0.19 and -0.19 \AA, respectively, and consequently it still has the inversion symmetry. On the contrary, one of the three Ti in $t$1T is shifted upward by 0.28 \AA, while the remaining two are shifted downward by 0.14 \AA. Such asymmetric shifts break spatial inversion symmetry and produce a separation of positive and negative charge centers. As a result, $t$1T shows a spontaneous polarization of 0.5 $\mu$C/cm$^2$ along the out-of-plane direction, but zero in-plane polarization. The polarization strength is nearly 1.4 times that in MoS$_2$\cite{Shirodkar,note}. Presence of out-of-plane ferroelectricity would lead to a kink in the corresponding potential energy profile\cite{WangJPCL}, which is indeed the case here. The shift of Ti changes the Ti-O bonds from a uniform 1.98 \AA\ in 1T to 1.87$\sim$2.12 \AA\ in $d$1T or $t$1T, with the Ti-Ti spacing narrows or elongates by 0.02$\sim$0.07 \AA.

For applications, out-of-plane ferroelectricity is more advantageous than in-plane ferroelectricity, yet out-of-plane ferroelectricity tends to disappear below a critical thickness due to depolarizing field effects\cite{Shang}. In contrast, in-plane polarization is insensitive to thickness and remains stable at the atomic thickness limit\cite{Chang}. Out-of-plane ferroelectricity is first theoretically predicted in monolayer MoS$_2$\cite{Shirodkar} and experimentally verified five years later in MoTe$_2$\cite{Yuan}. Interestingly, similar distorted 1$\times \sqrt{3}$ and $\sqrt{3}$$\times$$\sqrt{3}$ structures exist for MoS$_2$/MoTe$_2$, also the former centrosymmetric and the latter ferroelectric\cite{Bruyer}. In view of this, a comparison with them helps to deepen the understanding of structural instability and ferroelectricity observed here.

Under the 1T configuration, the local $D_{\rm {3d}}$ field splits the metal 3$d$-orbitals into a lower-lying $t_{\rm {2g}}$ triplet and a higher-lying $e_{\rm g}$ doublet. For the Mo of group 6, after crystallisation into MoS$_2$/MoTe$_2$, the two remaining $d$ electrons can only partially occupy the $t_{\rm {2g}}$ triplet. This triggers the Jahn-Teller instability, leading to a further splitting of the $t_{\rm {2g}}$ triplet, accompanied by a metal-insulator transition. Differently, the Ti belongs to group 4, so the characteristic electronic configuration of TiO$_2$ is always $d^0$, regardless of its crystal structure. Therefore, its 3$d$ occupancy must be empty, with the result that each configuration of TiO$_2$ has a bandgap separating the O-2$p$ from Ti-3$d$ states. This is evidenced by our first-principles calculations shown later.

Hence there must be another mechanism driving the deformation of 1T-TiO$_2$. One possibility is excitonic insulators, i.e., lattice distortion is triggered when electrons and holes at different $k$ points in the Brillouin zone spontaneously form excitons due to Coulomb attraction\cite{6,7,8,9}. While this mechanism has traditionally been introduced for semimetals with small band overlaps or semiconductors with small bandgaps, in recent years it has been found that the excitonic instability may also play a role in atomic-thick monolayers with one-electron gaps larger than 3 eV\cite{8,9}. To this end, we have calculated the exciton spectrum of 1T using Yambo code\cite{yambo2019}, and found a minimum exciton formation energy of about 3 eV, thus ruling out the excitonic insulator mechanism.

\begin{figure}[htbp]
\begin{center}
\includegraphics[width=0.9\columnwidth]{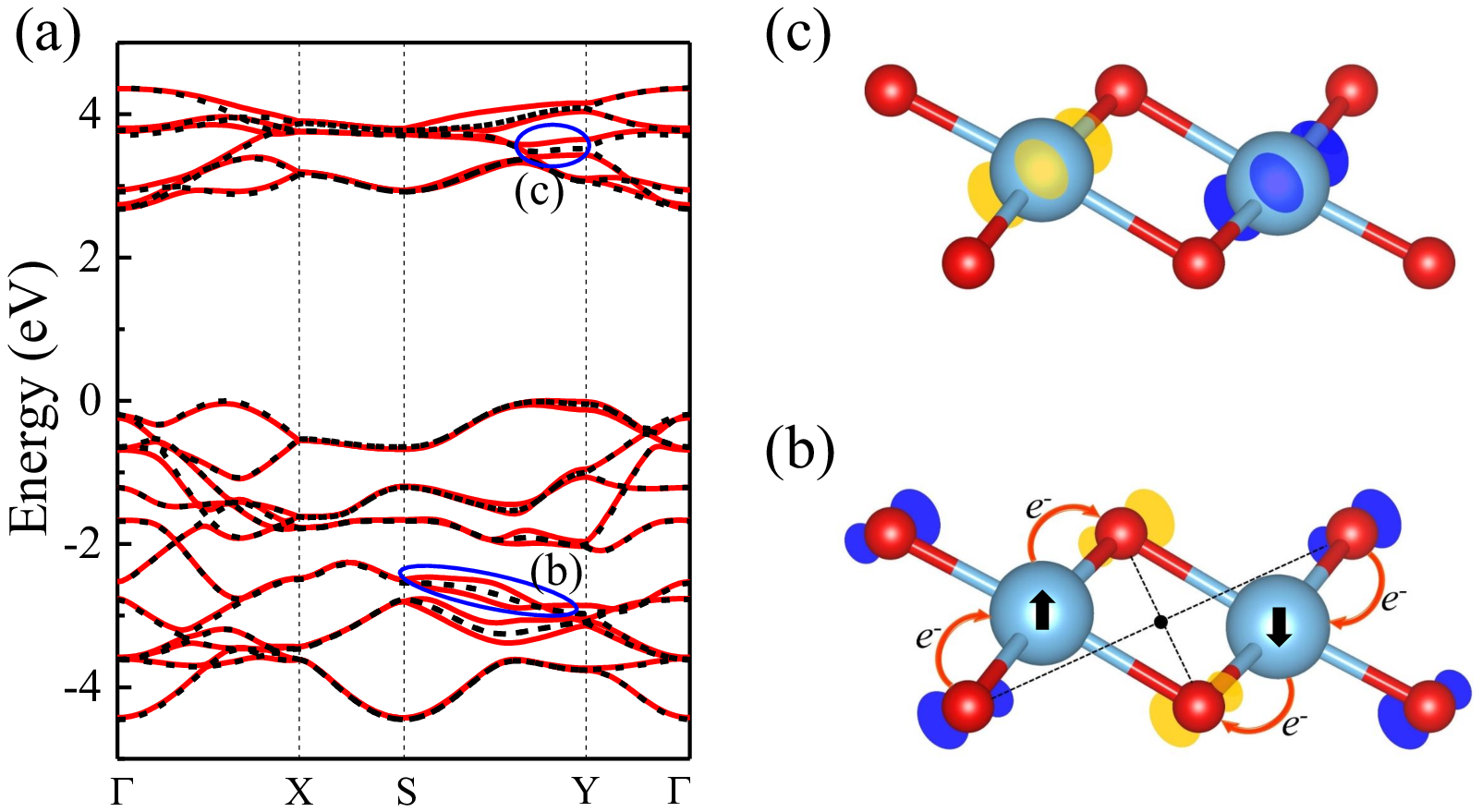}
\caption{\label{fig:fig2}(Color online) (a) Comparison between the PBE band structures of 1T (Black dashed lines) and second intermediate (Red solid lines). The valence band maximum is set as energy zero. (b)[(c)] Charge density difference between the two splitting O-$p$ (Ti-$d$) bands with an isosurface of 0.3 (0.1) $e$/\AA$^3$. The electron accumulation and depletion regions are depicted in yellow and blue, respectively. In (b), the red arrows illustrate the electron donation from the O-atom on one side to the empty Ti-$d$ orbitals and then back-donation to the O-atom on the other side. This drives the Ti-atom out of plane in the direction denoted by black arrows and makes it no longer the center of inversion. However, this does not destroy the other inversion center indicated by the black dot, so $d$1T is still centrosymmetric.}
\end{center}
\end{figure}

To understand the structural distortions here, we insert four intermediate structures uniformly between 1T and $d$1T. Figure 2(a) compares the bands of 1T and the second intermediate\cite{notefig2}, which reveals that the Ti displacement lifts the degeneracy of the occupied O-$p$ bands and the unoccupied Ti-$d$ bands along the $SY$ line. This hints at an interaction between the O-$p$ and Ti-$d$ bands and is a signal of the pseudo-Jahn-Teller effect\cite{pjet,Bersuker}. In Figs. 2(b) and 2(c), we representatively plot the charge density difference between the two splitting occupied and unoccupied bands circled in blue. As can be seen from Fig. 2(b), the pseudo-Jahn-Teller interaction leads to a charge redistribution between the upper-layer and lower-layer O atoms connected to the same Ti, causing them to be no longer equivalent. This breaks the inversion symmetry with respect to the Ti atoms. Bader analysis shows that the O atoms in the blue (yellow) regions of Fig. 2(b) lose (gain) a small number of electrons. Combining the above facts we can build a picture of the structural distortion due to the pseudo-Jahn-Teller interaction. It triggers a charge transfer in which electrons are donated from one O atom to the empty Ti $d$-orbitals and then back-donated to another O atom as depicted by the red arrows in Fig. 2(b). This donation-back donation drives the central Ti-atom to move vertically in the direction of charge transfer, reducing the system energy while inducing structural distortions. However, we note that the other inversion center of 1T [black dot in Fig. 2(b)] is preserved under the pseudo-Jahn-Teller interaction. Its existence forces the two Ti-atoms to move symmetrically up and down, corresponding to the split higher and lower conduction band [see Fig. 2(c)]. The case of $t$1T is quite different, because there is no center of symmetry between the three Ti. So their disproportionate movements in the out-of-plane direction lead to spontaneous polarization.

We further consider the phonon softening condition of the pseudo-Jahn-Teller theory\cite{Bersuker}
\begin{equation} \label{eq1}
\langle\Delta\rangle \textless \frac{8F^2}{K_{0}}.
\end{equation}
The $\langle\Delta\rangle$ is half of the energy gap between the mixing electronic states, which is derived directly from the band structure of 1T. The $F$ and $K_{0}$ are the vibronic coupling constant and the primary force constant, respectively, which can be fitted from the energetic calculations of the DFT. Within the pseudo-Jahn-Teller theory, the adiabatic potential energy surface $U$($Q$) is expressed as a function of the structural displacement $Q$ ($Q^2=Q_x^2 + Q_y^2 + Q_z^2$)
\begin{equation} \label{eq2}
U(Q) = \frac{1}{2}K_{0}Q^2-2\sum_{a=x,y,z} [\sqrt{\Delta^2+2F^2(Q^2-Q_a^2)}].
\end{equation}
Take for example the calculation of the $K$ point where phonon softening occurs (see Fig. 1). From the band structure of 1T, we have $\langle\Delta\rangle$ = 1.97 eV. Then, using DFT total energies of 1T, the second intermediate, and $d$1T, and their $Q$ displacements with respect to 1T, we get $F$ = 1.7 eV/\AA, $K_{0}$ = 11 eV/\AA$^2$, and $8F^2$/$K_0$ = 2.1 eV. Unambiguously, Eq. (1) is satisfied, justifying that it is indeed the pseudo-Jahn-Teller mechanism at work.

\begin{table}
\centering
\caption{Summarization of the space group, the minimum gaps calculated by different methods, and optical gap ($E_{\rm opt}$, defined as the first peak on the imaginary part of the calculated dielectric function.) of various two-dimensional TiO$_2$ monolayers. The bulk rutile is taken for a comparison.}
\renewcommand\arraystretch{1.0}
\begin{ruledtabular}
\begin{tabular}{lcccccccccccccccccccccccccc}
\multirow{2}{*}{TiO$_2$} & \multirow{2}{*}{Space group} & \multicolumn{3}{c}{Minimum gap (eV)}   & \multirow{2}{*}{$E_{\rm opt}$ (eV)}  \\
  \cline{3-5}
                   \multirow{2}{*}&         & PBE  &HSE06 &$G_{0}W_{0}$ &  \\
\hline
           $t$1T    &{$P31m$}               &2.92  &4.52  &5.38 &3.99\\
           $d$1T    &{$P2_1/m$}             &2.85  &4.45  &5.36 &4.21\\
           1T       &{$P\overline{3}m1$}    &2.67  &4.12  &5.09 &3.68\\
           1H       &{$P\overline{6}m2$}    &1.18  &2.61  &3.74 &3.69\\
           rutile   &{$P4_2/mnm$}           &1.82  &3.32  &3.22 &3.94\\
\end{tabular}
\end{ruledtabular}
\end{table}

Next, we turn to the electronic properties. Since PBE severely underestimates the bandgap of bulk TiO$_2$, we carry out different levels of calculations. Rutile TiO$_2$, which is the most stable bulk phase, is used for comparison. Some of the characteristic results are listed in Table I and the typical band structures of $d$1T and $t$1T are plotted in Fig. 3. Rutile-TiO$_2$ has a PBE bandgap of 1.82 eV and an optical bandgap of 3.94 eV, which are in agreement with previous studies\cite{ZhuJPCC}. Although the calculated bandgaps are method-dependent, both the fundamental and optical bandgaps of $d$1T and $t$1T are significantly larger than those of rutile under the same method. In fact, the fundamental bandgap obtained by each method for 1T is also larger than that of rutile. Nevertheless, the PBE and HSE06 bandgaps of 1H are smaller than that of rutile. Comparison of these monolayer results reveals a tendency for the fundamental bandgap to increase as the system energy decreases.

Our preceding structural calculations indicate that the freestanding TiO$_2$ monolayer does exist. Whether it is 1T, $d$1T, or $t$1T, it matches the experimentally observed planar honeycomb-like hexagonal crystal structure and the measured thickness\cite{NM2021}. However, as far as the bandgap is concerned, our comparisons at the same level consistently support the fact that the bandgap of the monolayer is larger, not smaller, than that of the bulk. One aspect of the bandgap discrepancy may arise from first-principles calculations. It is well known that there is no method that ensures reproducing the correct bandgap. Even the method that reproduces the bulk TiO$_2$ bandgap is not guaranteed to yield the correct bandgap for the monolayer. On the other hand, the experimental conditions are different from first-principles simulations in terms of temperature, substrate and sample (e.g., defects). All these may lead to variations in the bandgap.

\begin{figure}[htbp]
\begin{center}
\includegraphics[width=0.9\columnwidth]{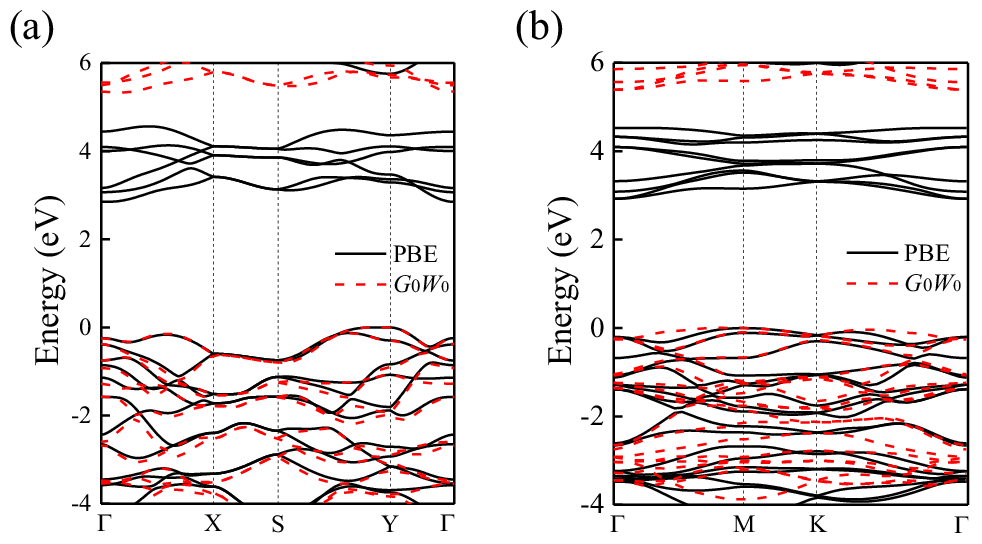}
\caption{\label{fig:fig2}(Color online) Band structures of (a) $d$1T-TiO$_2$ and (b) $t$1T-TiO$_2$, respectively calculated by PBE and G$_0$W$_0$. The valence band maximum is set as energy zero.}
\end{center}
\end{figure}

Here we briefly discuss some experimental possibilities. One is the strain effect of the substrate used in the experiment, which does reduce the bandgap. For example, a 5\% in-plane strain shrinks the PBE bandgap of $d$1T and $t$1T by about 5\% and 11\%, to 2.72 and 2.60 eV, respectively. From this estimation, a strain of at least 17\% is required to make the fundamental bandgap smaller than that of the bulk, which sounds difficult. Nonetheless, their optical bandgaps of 3.80 and 3.60 eV do become smaller compared to the bulk rutile.

Another possibility is the presence of 1H due to temperature-enhanced stability, which have relatively small bandgaps (see Table I). In order to calculate the effect of temperature on the phonon spectrum, we vary the Fermi-Dirac distribution function that describes the probability of occupation of the electronic states by varying the smearing parameter. By this means, we can assess the structural stability at different temperatures\cite{prb1965}. It was found that the soft phonon modes of both 1H and 1T gradually harden as the smearing increases. At a smearing of 0.8 eV, there are no more soft modes on the phonon spectrum (not shown). Although smearing does not directly correspond to the temperature, such dependence suggests that 1H and 1T can stabilize at sufficiently high temperatures. However, on the other hand, our energetic calculations show that the relative stability trend of $d$1T $>$ 1T $>$ 1H remains unchanged even in the presence of strain ($\leq$ $\pm$5\%) or electronic doping ($\leq$ $2.5\times10^{14}$/cm$^2$). These point to a crucial role of kinetic factors and/or entropic effects in the experimental observations.

In summary, by first-principles calculations, we predict for the first time stable hexagonal monolayers of $d$1T-TiO$_2$ and $t$1T-TiO$_2$, the latter of which is particularly intriguing because of its intrinsic out-of-plane ferroelectricity. It is the $d^0$ nature of Ti that distinguishes the physics associated with structural distortion here from the widely studied $d^2$ dichalcogenides. First, it occurs in a wide-gap semiconductor with the absence of a metal-insulator transition. Second, it is driven by a pseudo-Jahn-Teller effect involving interactions with excited states. Third, it is characterized by out-of-plane motions of metal atoms instead of in-plane dimerization and trimerization. These results provide useful insights into the understanding of the recent experiment. On the other hand, it is interesting to note that the $d^0$ and $d^2$ systems exhibit the commonality of both having two distorted 1T structures, one of which remains centrosymmetric while the other is ferroelectric. Whether this is a commonality dictated by the two centers of symmetry of the 1T structures is worth exploring in the future.

This work was supported by the Ministry of Science and Technology of China (Grant Nos. 2023YFA1406400 and 2020YFA0308800) and the National Natural Science Foundation of China (Grant Nos. 12074034 and 11874089).


\begin{thebibliography}{90}
\bibitem{note} Since the bulk TiO$_2$ has a non-layered structure, we calculate the polarisation strength here using the thickness of a monolayer, i.e., 3 and 2 \AA\ for MoS$_2$ and $t$1T-TiO$_2$, respectively. Thus, the result of 0.36 $\mu$C/cm$^2$ differs from 0.18 $\mu$C/cm$^2$ in Ref. \onlinecite{Shirodkar}.
\bibitem{notefig2} Here physics does not depend on which intermediate is chosen, just for clarity.
\end{thebibliography}
\end{document}